\documentclass[reqno,pdftex]{aastex631}

\usepackage{amsmath}
\usepackage{needspace}
\usepackage{float}

\newcommand{\chandra}{Chandra}
\newcommand{\xmm}{XMM-Newton}
\newcommand{\nustar}{NuSTAR}
\begin{document}

\title{A Chandra X-Ray Survey of the Middle-aged Supernova Remnant G22.7$-$0.2 Interacting with Molecular Clouds}

\author[0009-0002-1660-2420]{Harry Van Der Ark}
\author[0000-0002-9709-5389]{Kaya Mori}
\author{Charles J. Hailey}
\affiliation{Columbia Astrophysics Laboratory, Columbia University, New York, NY 10027, USA}

\begin{abstract}
G22.7$-$0.2 is a middle-aged supernova remnant (SNR; $\tau \sim 3\times10^4$~yr) interacting with the nearby giant molecular cloud GMC~23.3$-$0.4. A handful of X-ray sources have been identified within the SNR radio shell by previous \xmm\ and \nustar\ observations. In particular, three of those X-ray sources (N1, N2, and N3) have been detected above 10 keV, indicating that they are either X-ray point sources, or molecular cloud clumps heated by SNR blast waves. In the latter case, non-thermal bremsstrahlung or synchrotron radiation from shock-heated molecular clouds may appear as extended X-ray sources. We conducted an X-ray survey of the SNR-cloud interaction region with \chandra\ and detected 23 X-ray sources with $> 5\sigma$ significance using {\tt wavdetect}, 20 of which were previously uncataloged. The sub-arcsecond angular resolution of \chandra\ allowed us to spatially resolve these X-ray sources. After accounting for point spread function (PSF) smearing and dust scattering effects, we determine that two \nustar\ hard X-ray sources (N1 and N2) are extended, whereas the other \chandra\ sources are point sources. While N1 may be a compact filament or a synchrotron-emitting clump with $r\sim0.018$~pc, our analysis suggests that N2 is likely a compact cloud clump shock-heated by the SNR and emitting non-thermal bremsstrahlung or synchrotron X-rays, with $r\sim0.052$~pc and $M\approx0.6\,M_\odot$ under the non-thermal bremsstrahlung model. 
\end{abstract}

\keywords{Supernova remnants (1667) --- Molecular clouds (1072) --- X-ray sources (1822) --- Non-thermal radiation sources (1119) --- Interstellar medium (847)}

\section{Introduction}\label{sec:intro}
Molecular cloud (MC) substructures include clumps, cores, and filaments; clumps in particular are confined, dense clouds with velocity coherence. Their masses and physical sizes are described by the clump mass function (CMF)  $dN/d\ln M \propto M^{-\alpha}$. The CMF is constrained well at higher masses, but is poorly constrained around the theoretical minimum clump size $r \sim 0.01$--$0.1$~pc \citep{1998ApJ...497..842P,2007ARA&A..45..565M,Lombardi_2010}. A possible break in the CMF at this regime could suggest a change in MC formation physics that could affect star formation, since a similar break was observed in the initial mass function \citep{2007ARA&A..45..565M}. In fact, $r \sim 0.01$--$0.1$~pc is considered the critical length scale for cloud formation and fragmentation both theoretically and observationally \citep{10.1093/mnras/staa1262}. Thus, compact cloud clumps (CCCs) with size $r \sim 0.01$--$0.1$~pc offer a unique testbed for exploring gas dynamics and star formation.  

Compact ($r\sim0.01$--$0.1$ pc) X-ray structures associated with individual MC clumps have been reported across a range of supernova remnant (SNR) ages, but a non-thermal origin has so far only been established for those around young SNRs, such as  RX J1713.7$-$3946 ($\tau\sim2\times10^3$ yr), whose ALMA-resolved sizes are consistent with CCCs \citep{2020ApJ...904L..24S}. There, \citet{2020ApJ...899..102H} and \citet{Kawabata2026upstreamB} attribute the hotspots to synchrotron radiation from primary electrons accelerated at the forward shock, or from secondary $e^\pm$ produced by hadronic (pp) collisions between shock-accelerated protons and dense clump gas, both radiating in a locally amplified B-field where the shock compresses dense material \citep{2012ApJ...744...71I}. Similarly sized knots identified in the middle-aged SNR W44 ($\tau\sim2\times10^4$ yr) were thought to be attributable to MC evaporation \citep{1991ApJ...373..543W}, but could only be modeled with thermal emission models \citep{Shelton_2004}. No non-thermal spectral component has yet been established for a structure ($r\sim0.01$--$0.1$ pc) at these ages. This is consistent with the expected decline of the accelerated lepton populations with SNR age: \citet{2012ApJ...746..134N} find that synchrotron X-ray luminosity drops sharply once the shock has decelerated enough that the maximum electron energy, and hence the synchrotron frequency, falls below the X-ray band. Because primary-electron hotspots draw on the same shock-accelerated electron population as the shell, their maximum synchrotron energy should decline on the same timescale as the shell's; secondary-electron hotspots instead track the maximum energy of the parent protons, which declines more gradually with age but under the same shock deceleration \citep{2012ApJ...746..134N}.

Over 25 years ago, it was predicted that a SNR shell colliding with molecular clouds can accelerate electrons via shocks, producing low-energy cosmic rays (LECRs; keV--MeV electrons) that bombard the clouds and generate non-thermal bremsstrahlung (NTB) radiation \citep{Bykov_2000}. This NTB emission emerges in the X-ray band from sub-GeV electrons and can persist in SNRs up to $\tau\sim10^5$ yr. The CCC synchrotron scenario described above \citep{2020ApJ...899..102H,Kawabata2026upstreamB} is not excluded at these ages, but becomes harder to sustain as the shock decelerates and the maximum lepton energy declines \citep{2012ApJ...746..134N}. It also requires core-scale densities ($n\sim10^5$--$10^7\,{\rm cm}^{-3}$) that grow increasingly rare in this regime, needed to supply both B-field amplification and proton penetration \citep{2020ApJ...899..102H}. NTB radiation, by contrast, requires only modestly accelerated (sub-GeV) electrons and far more common clump densities, making it the more readily sustained channel in older, middle-aged SNRs ($\tau\sim10^4$--$10^5$ yr). Observationally confirming NTB-emitting CCCs within a middle-aged SNR would establish an additional model for CCC X-ray illumination and significantly expand the search criteria for SNRs hosting X-ray-emitting CCCs.

\begin{figure}
    \centering
    \includegraphics[width=0.75\linewidth]{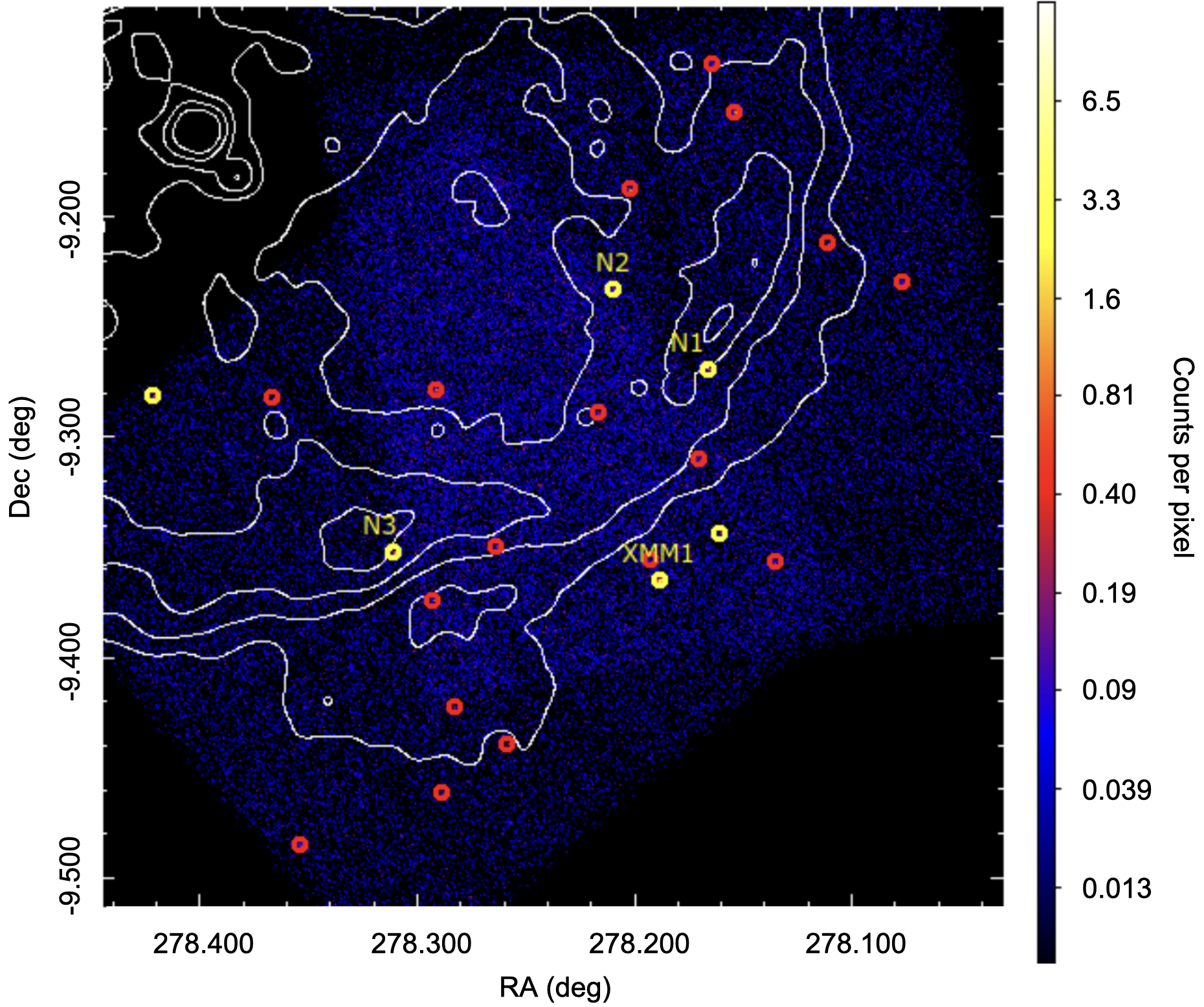}
    \caption{Merged and smoothed \chandra\, observations overlaid with Multi-Array Galactic Plane Imaging Survey (MAGPIS) 20 cm radio contours \citep{Helfand_2006} of the G22.7$-$0.2 SNR shell (white). Six sources detected in previous XMM+\nustar\, observations are circled in yellow, including the known X-ray binary XMM1 and CCC candidates N1, N2, and N3. Seventeen newly detected \chandra\ sources from this survey with significance $>5\sigma$ are circled in red. Axes denote equatorial RA and Dec in degrees.}
    \label{fig:merged}
\end{figure}

The diagnostic signature of NTB energy loss to clouds is a flat X-ray spectrum with photon index $\Gamma\simeq1$ \citep{Bykov_2000,Uchiyama_2002,10.1111/j.1365-2966.2009.14832.x}. This is well toward the hard end of the broader $\Gamma\sim0.5$--$6$ range spanned by the individual synchrotron-attributed hotspots \citep{2020ApJ...899..102H}. To date, no clear evidence for the flat NTB spectrum of a CCC has been found at a SNR--MC interaction site, other than large filament structures, because observations have been limited by dominant synchrotron emission in younger remnants \citep{2015ApJ...800..119N}. The best candidate site identified so far is the shell-type remnant G22.7$-$0.2, at a distance of 4.4--4.7\,kpc, from the kinematic association with $\sim$75--79\,km\,s$^{-1}$ molecular clouds showing shock-broadened CO line wings \citep{Su_2014} and from HI absorption against the shell \citep{Ranasinghe_2018}. Age was constrained to $\tau\gtrsim3\times10^{4}$\,yr from the XMM-Newton shock speed measurement $\lesssim$200\,km\,s$^{-1}$ for the SNR scale $r\sim18$\,pc \citep{li2026}, consistent with the $(2.5$--$5)\times10^{4}$\,yr detectable lifetime of SNRs \citep{Helfand_2006}. Millimeter CO observations show its shell interacting with nearby MCs of GMC~23.3$-$0.4 \citep{Su_2014}, at a hydrogen density of $\sim5.3\times10^{2}\,{\rm cm}^{-3}$ \citep{2026ApJ..1001...19M}, well within the regime for which the NTB model was originally demonstrated \citep{Bykov_2000}. However, it is several orders of magnitude below the core-scale densities the secondary-electron synchrotron mechanism requires, making a synchrotron origin comparatively disfavored here. The region contains a bright X-ray source (XMM~J183245$-$0921539, hereafter XMM1), the counterpart to the $\gamma$-ray binary HESS~J1832$-$093 (Figure~\ref{fig:merged}) \citep{10.1093/mnras/stu2148}.

As shown in Figure~\ref{fig:merged}, combined XMM+\nustar\, observations detected a handful of discrete X-ray sources aligned with the shell \citep{Mori_2017}. Apart from XMM1, the only sources detected by \nustar\ with emission visible at 10--20~keV are N1, N2, and N3. Because the NTB and Synchrotron signature is a flat spectrum extending to hard X-rays, these are the only relevant sources to test. Their joint XMM+\nustar\, fits give $\Gamma = 2.1^{+0.5}_{-0.4}$ for N1 (too soft for NTB, possibly synchrotron), $\Gamma = 1.2^{+0.4}_{-0.4}$ for N2, and $\Gamma = 0.9^{+0.4}_{-0.3}$ for N3 \citep{Mori_2017}, so N2 and N3 are the leading NTB candidates. Using XMM and \nustar\, data alone, \citet{Mori_2017} proposed that N1 and N2 are possible X-ray binaries or magnetic cataclysmic variables, and that N3 may be compact diffuse emission from a supernova-ejecta fragment interacting with a dense cloud, as seen in IC~443 \citep{Bykov_2005}; they noted that follow-up \chandra\, observations were needed to resolve the sources' spatial extent and identify infrared counterparts. Although neutron star high-mass X-ray binaries (HMXBs) can also exhibit flat X-ray spectra, this is unlikely given the absence of bright nearby infrared counterparts ($K < 15$, \S\ref{sec:ir}) and X-ray luminosities ($L_X \sim 6 \times 10^{32}$~erg~s$^{-1}$, \citep{Mori_2017}) too low for HMXBs. Conversely, N1's softer spectrum and proximity to a region of enhanced radio flux (contours in Figure~\ref{fig:merged}) suggest it may be a filament with locally amplified magnetic fields emitting synchrotron X-rays \citep{Uchiyama_2008,Zhang_2014,Kawabata2026upstreamB}. However, none of these sources can be spatially resolved as extended CCCs with \nustar\, at 10--20~keV ($\Delta\theta \simeq 18''$ FWHM, corresponding to $r \simeq 0.4$~pc).

Given the multi-wavelength observations of G22.7$-$0.2 at the SNR--MC interaction front, N2, N3, and potentially N1 are strong candidates for compact diffuse X-ray-emitting CCCs. \chandra's sub-arcsecond resolution in the 2--8~keV band ($\Delta\theta \simeq 0.5''$ FWHM on-axis, corresponding to $r \simeq 0.01$~pc) enables spatial resolution at the scale where the smallest CCCs are predicted to form. In 2024, \chandra\ targeted the southern boundary of G22.7$-$0.2, covering N1, N2, and N3, to measure their spatial extents. Confirming these X-ray-emitting CCCs would provide a novel observational tool for probing molecular cloud fragmentation and formation physics at sub-parsec scales.

This paper is organized as follows. Section~\ref{sec:obs} describes the \chandra\ observations, source detection methods (\S\ref{sec:chandra_obs}), and cross-matching with the 2MASS infrared catalog to search for IR counterparts in case they are X-ray Binaries (\S\ref{sec:ir}). Section~\ref{sec:imaging} details the spatial analysis methods used to resolve source extent, including PSF modeling (\S\ref{sec:psf}), radial profile fitting (\S\ref{sec:radial}), and PSF deconvolution (\S\ref{sec:deconv}). Section~\ref{sec:dust} evaluates potential dust-scattering contributions. Section~\ref{sec:results} evaluates the physical nature of candidates N1--N3 (\S\ref{sec:cand}), and discusses the broader implications and future directions of this work (\S\ref{sec:implications}). We summarize our conclusions in \S\ref{sec:conclusion} and show the source versus PSF images for 23 documented sources in Appendix~\ref{app:psf}.

\section{Observations and Source Cataloging}\label{sec:obs}
For this work we make use of data from the \chandra\ X-ray Observatory and the Two Micron All Sky Survey (2MASS). \chandra\ has been NASA's flagship X-ray mission since its launch in 1999, with an effective energy range of 0.1--10 keV \citep{Weisskopf_2002}. Its High Resolution Mirror Assembly (HRMA) gives the ACIS-I instrument the highest on-axis angular resolution of any X-ray telescope yet developed ($\sim 0.5''$), making it uniquely suited for this work. 2MASS is a ground-based all-sky survey that detects infrared sources in the $J$, $H$, and $K_s$ bands with angular resolution $\sim2.9''$ in the $K_s$ band \citep{2006AJ....131.1163S}. Its uniform coverage of G22.7$-$0.2 makes it suitable for detecting IR counterparts to X-ray sources. In this section we present our \chandra\ observations targeting the southern boundary of G22.7$-$0.2, describe our source detection methods, and describe our search for infrared counterparts in the 2MASS catalog to assess multi-wavelength associations.

\subsection{Chandra X-Ray Observations and Source Detection}\label{sec:chandra_obs}
The data consist of four \chandra\ Advanced CCD Imaging Spectrometer (ACIS-I) observations (ObsIDs: 26681, 26682, 29265, 29266) made in 2024, covering the X-ray sources along the southern boundary of the G22.7$-$0.2 shell (Figure~\ref{fig:merged}).  Pointings targeting N1 (29.73~ks) and N3 (28.32~ks) position the three CCC candidates near the center of the field of view to minimize off-axis point spread function (PSF) degradation and optimize spatial resolution. These observations yield 58.05~ks of total exposure time over an area of $\sim510$~arcmin$^2$, with $\sim430$~arcmin$^2$ receiving $\gtrsim30$~ks exposure. Images of the hard sources XMM1, N1, N2, and N3 were made in the 2--8~keV band, where these sources emit most of their detectable flux.

As a secondary objective, we searched for additional prominent X-ray sources in the 0.5--7.0~keV energy band using {\tt wavdetect}, the source detection tool included with the \chandra\ Interactive Analysis of Observations (CIAO) software (version 4.17, with CALDB 4.12.4) \citep{Freeman_2002, 2006SPIE.6270E..1VF}. We set detection thresholds to be inclusive, subsequently filtering out sources with significance $<5\sigma$. This yielded 23 significant sources, six of which were previously detected by XMM+\nustar. Of these six, XMM1 and two active galactic nuclei (AGNs) are previously cataloged X-ray sources, whereas N1, N2, and N3 have been discussed in \citet{Mori_2017} but do not appear in a public X-ray catalog. This resulted in 20 sources with no previous X-ray catalog entry (including N1, N2, and N3). Key output parameters from {\tt wavdetect}, along with absorbed flux estimates, and 2MASS counterparts from \S\ref{sec:ir} are listed in Table~\ref{tab:xray_sources}.

Many previous studies of SNR--MC interactions have been limited by low angular resolution and the faintness of MC substructures in the IR, sub-mm, and radio bands \citep{Su_2014, 2026ApJ..1001...19M}. The \chandra\, survey of this region is the first to resolve these problems and produce a usable catalog. Figure~\ref{fig:merged} shows these sources in the context of the G22.7$-$0.2 SNR and merged \chandra\, observations. Along the southern boundary covered by these observations, we detect no significant diffuse X-ray emission from the shell in the 0.5--7.0~keV band, consistent with the remnant's advanced age: at a shock speed of $\lesssim$200~km~s$^{-1}$ \citep{li2026}, the post-shock temperature is $kT \lesssim 0.06$~keV, too cool to produce diffuse thermal emission in the \chandra\ band \citep{Zhekov_2005}, and the shock is too slow to accelerate electrons to the energies needed for synchrotron X-rays \citep{2012ApJ...746..134N}. Notably, N1, N2, and N3 lie within the SNR radio contours, whereas XMM1 lies in front of them. This makes XMM1 a useful control for the methodology applied to N1, N2, and N3 in \S\ref{sec:imaging}. 

\begin{longrotatetable}
\begin{deluxetable*}{lccccccccc}
\setlength{\tabcolsep}{4pt}
\tabletypesize{\footnotesize}
\tablecaption{Significant X-Ray Sources Detected in the Chandra Broad Band\label{tab:xray_sources}}
\tablehead{
\colhead{} & \colhead{} & \colhead{} & \colhead{2$\sigma$ R.A.} & \colhead{2$\sigma$ Decl.} & \colhead{Net} & \colhead{} & \colhead{Mean} & \colhead{Absorbed Flux} & \colhead{2MASS} \\
\colhead{Source Name} & \colhead{R.A.} & \colhead{Decl.} & \colhead{Uncertainty} & \colhead{Uncertainty} & \colhead{Counts} & \colhead{Significance} & \colhead{Off-Axis Angle} & \colhead{Estimate} & \colhead{Counterpart} \\
\colhead{} & \colhead{(deg)} & \colhead{(deg)} & \colhead{(arcsec)} & \colhead{(arcsec)} & \colhead{} & \colhead{($\sigma$)} & \colhead{(arcmin)} & \colhead{(10$^{-14}$} & \colhead{} \\
\colhead{} & \colhead{} & \colhead{} & \colhead{} & \colhead{} & \colhead{} & \colhead{} & \colhead{} & \colhead{erg cm$^{-2}$ s$^{-1}$)} & \colhead{}
}
\startdata
XMM J183245$-$0921539 & 278.18811 & $-$9.36532 & 0.256 & 0.204 & 447 & 89.0 & 6.39 & 20.0 & 2MASS J18324516$-$0921545 \\
CXO J183250$-$091401 (N2) & 278.20901 & $-$9.23367 & 0.170 & 0.202 & 113 & 42.0 & 2.82 & 6.36 & \nodata \\
CXO J183248$-$091117 & 278.20183 & $-$9.18790 & 0.284 & 0.242 & 100 & 41.0 & 3.43 & 7.22 & \nodata \\
CXO J183237$-$090909 & 278.15346 & $-$9.15271 & 0.304 & 0.300 & 125 & 38.0 & 5.69 & 9.08 & \nodata \\
CXO J183314$-$092110 (N3) & 278.31019 & $-$9.35267 & 0.159 & 0.137 & 57 & 27.0 & 0.39 & 3.19 & \nodata \\
CXO J183240$-$091610 (N1) & 278.16573 & $-$9.26960 & 0.137 & 0.147 & 57 & 27.0 & 1.94 & 4.16 & \nodata \\
CXO J183309$-$091643 & 278.29093 & $-$9.27875 & 0.524 & 0.430 & 90 & 23.0 & 6.64 & 3.33 & \nodata \\
CXO J183226$-$091246 & 278.11070 & $-$9.21269 & 0.414 & 0.374 & 28 & 16.0 & 4.88 & 2.89 & \nodata \\
CXO J183239$-$090753 & 278.16360 & $-$9.13138 & 0.896 & 0.788 & 54 & 15.0 & 6.71 & 4.08 & \nodata \\
XMM J183340.8$-$091653 & 278.42107 & $-$9.28193 & 0.856 & 1.002 & 53 & 14.0 & 8.17 & 4.05 & \nodata \\
CXO J183325$-$092908 & 278.35361 & $-$9.48557 & 1.132 & 1.118 & 43 & 11.0 & 8.46 & 3.31 & \nodata \\
CXO J183302$-$092623 & 278.25776 & $-$9.43973 & 0.714 & 0.576 & 21 & 8.4 & 5.84 & 1.57 & \nodata \\
CXO J183327$-$091658 & 278.36613 & $-$9.28276 & 0.546 & 0.634 & 20 & 8.2 & 5.63 & 1.48 & 2MASS J18332786$-$0916577 \\
CXO J183218$-$091349 & 278.07648 & $-$9.23028 & 0.626 & 0.862 & 23 & 7.6 & 6.44 & 1.65 & 2MASS J18321822$-$0913484 \\
CXO J183241$-$091838 & 278.16974 & $-$9.31072 & 0.844 & 1.032 & 25 & 6.9 & 4.28 & 1.12 & 2MASS J18324069$-$0918372 \\
CXO J183310$-$092228 & 278.29221 & $-$9.37442 & 0.672 & 0.314 & 18 & 6.6 & 1.43 & 0.80 & \nodata \\
CXO J183307$-$092523 & 278.28250 & $-$9.42290 & 0.684 & 0.310 & 14 & 6.4 & 4.35 & 1.09 & \nodata \\
CXO J183309$-$092744 & 278.28792 & $-$9.46218 & 0.894 & 0.864 & 17 & 6.2 & 5.99 & 1.28 & \nodata \\
CXO J183245$-$092124 & 278.19191 & $-$9.35659 & 0.974 & 0.856 & 20 & 6.2 & 6.59 & 0.89 & \nodata \\
CXO J183252$-$091723 & 278.21626 & $-$9.28959 & 0.502 & 0.636 & 16 & 5.8 & 4.66 & 0.71 & \nodata \\
XMM J183238$-$092016 & 278.16067 & $-$9.33778 & 1.288 & 0.992 & 19 & 5.7 & 4.82 & 1.08 & \nodata \\
CXO J183303$-$092100 & 278.26328 & $-$9.35012 & 0.588 & 0.514 & 16 & 5.7 & 5.83 & 0.58 & 2MASS J18330319$-$0921003 \\
CXO J183232$-$092126 & 278.13478 & $-$9.35725 & 1.758 & 0.988 & 23 & 5.5 & 5.95 & 1.04 & \nodata \\
\enddata
\tablecomments{Sources are ordered by significance. The broad band is 0.5--7.0 keV. R.A. and decl. uncertainties denote 2$\sigma$ positional uncertainties from the {\tt wavdetect} centroid function. Net counts represent source counts after subtracting local background. Absorbed flux estimates (0.5--7.0 keV) for newly detected sources were calculated using PIMMS \citep{Mukai1993PIMMS}, assuming a $\Gamma=1.7$ power-law spectrum. We designate N1, N2, and N3 using standard \chandra\ naming conventions for their first public catalog. Two sources, XMM J183340.8$-$091653 and XMM J183238$-$092016, are previously cataloged AGNs observed by \xmm\ \citep{Webb2020_4XMM}.}
\end{deluxetable*}
\end{longrotatetable}

\subsection{Cross-matching with Infrared Sources}\label{sec:ir}
We investigated infrared (IR) counterparts to our X-ray detections, as cold dust in MCs and X-ray binary companions frequently produce IR emission \citep{2001ApJ...551..807D}. We cross-matched all \chandra\,  source positions against the 2MASS Point Source Catalog \citep{2MASS_PSC}, which provides complete coverage of the G22.7$-$0.2 field with $\sim20{,}000$ sources, detecting sources down to a limiting magnitude of $K_{\rm mag}\sim15.8$ in this field. Using the combined $2\sigma$ positional uncertainties of the \chandra\ sources (from {\tt wavdetect}) and the 2MASS sources as the search radii, five sources were identified to have 2MASS counterparts (Table~\ref{tab:xray_sources}). The only previously classified 2MASS counterpart was 2MASS J18324516$-$0921545, identified as the HMXB donor star to XMM1 \citep{Eger_2016}. These matches are not definitive, as chance alignments with foreground or background sources are possible. The absence of detected IR counterparts for N1, N2, and N3 supports their identification as CCCs, which are not expected to produce strong IR emission unlike some HMXBs and low-mass X-ray binaries. However, high line-of-sight extinction within GMC~23.3$-$0.4 could attenuate embedded IR emission, favoring the detection of foreground stars.

\section{Chandra Imaging Analysis}\label{sec:imaging}
Of the 23 sources in Table~\ref{tab:xray_sources}, we analyze the spatial extent of only N1, N2, and N3, with XMM1 serving as a control. These are the only sources detected by NuSTAR above 10~keV (\S\ref{sec:intro}), and therefore the only ones with non thermal hard X-ray emission expected of NTB or synchrotron emitting CCCs. To determine whether N1, N2, and N3 are extended CCCs, we compare each source to its PSF, which models the appearance of a point source after its light is reflected by \chandra's optics. If a source is more extended than its PSF, it is a probable CCC. We use the known point source XMM1 as a control, since it is expected to closely match its PSF. The method for distinguishing a point source from an extended source therefore begins with modeling the \chandra\,  PSF of each source. We then produce a normalized radial profile histogram for both the source and its PSF and compare their fitted models. If the source's radial profile deviates significantly from its PSF, we conclude that the source is extended. 

\subsection{Modeling the PSF}\label{sec:psf}
The \chandra\, PSF can be modeled using the \chandra\, Ray Tracer \citep[ChaRT v2;][]{2003ASPC..295..477C}, which simulates \chandra's optics by tracing rays through the HRMA to produce a rayfile. Another tool, the Model for AXAF Response to X-Rays (MARX, version 5.5.3), takes these rayfiles and produces a PSF image \citep{2012SPIE.8443E..1AD}. Running ChaRT requires the source position, telescope pointing, and the source spectrum. The \chandra\ observations of XMM1, N1, N2, and N3 had fewer spectral counts and higher uncertainty than the XMM+\nustar\ data; hence, we adopted the best-fit absorbed power-law models from the joint spectral fits of \citet{Mori_2017} as the input spectra: XMM+\nustar\, over 2--30~keV for XMM1, and XMM+\nustar\, over $\simeq$3--20~keV for N1, N2, and N3. We adopted the photon index, flux, energy ranges, and hydrogen column densities from \citet{Mori_2017} shown in Table~\ref{tab:chart_spectra}. These values were input into the PIMMS tool \citep{Mukai1993PIMMS} to calculate the corresponding \chandra\, count rate for flux normalization. We then combined the \chandra\,  flux normalization with the same spectral parameters, \chandra's energy range, the source's off-axis angle, and exposure to simulate the spectrum as observed by \chandra. Since the off-axis angle and exposure of a source vary between ObsIDs, a spectrum was produced for each source in each ObsID. We then ran ChaRT on each spectrum, processed the resulting rayfiles in MARX, and reprojected and merged them across ObsIDs to produce a complete PSF image for each source.

\begin{deluxetable}{lcccccc}
\tablecaption{Best-Fit Absorbed Power-Law Parameters Used as ChaRT Input Spectra from \citet{Mori_2017}\label{tab:chart_spectra}}
\tablehead{
\colhead{Source} & \colhead{Band} & \colhead{$N_{\rm H}$} & \colhead{$\Gamma$} & \colhead{$F_{\rm abs}$} & \colhead{$F_{\rm unabs}$} & \colhead{$\chi^2_\nu$ (dof)} \\
\colhead{} & \colhead{(keV)} & \colhead{($10^{22}$~cm$^{-2}$)} & \colhead{} & \colhead{(2--10~keV)} & \colhead{(2--20~keV)} & \colhead{}
}
\startdata
XMM1 & 2--30 & $9.5 \pm 2$ & $1.5 \pm 0.1$ & $8.0 \pm 0.8$ & \nodata & 1.0 (112) \\
N1 & $\sim$3--20 & $17^{+10}_{-7}$ & $2.1^{+0.5}_{-0.4}$ & \nodata & $3.7^{+1.5}_{-0.8}$ & 0.97 (34) \\
N2 & $\sim$3--20 & $11^{+7}_{-5}$ & $1.2 \pm 0.4$ & \nodata & $4.1^{+0.8}_{-0.6}$ & 0.62 (18) \\
N3 & $\sim$3--20 & $6^{+8}_{-4}$ & $0.9^{+0.4}_{-0.3}$ & \nodata & $4.2^{+0.7}_{-0.5}$ & 0.97 (28) \\
\enddata
\tablecomments{Flux is in units of $10^{-13}$~erg~cm$^{-2}$~s$^{-1}$. Uncertainties are at the 90\% confidence level.}
\end{deluxetable}

\needspace{5\baselineskip}
\subsection{Producing and Fitting the Radial Profiles}\label{sec:radial}
A radial profile is a histogram of a source's flux density at different radii. More extended sources have a radial profile in which the flux density is proportionately higher farther from the source center. To create a radial profile, annuli are constructed around the source with a defined center and bin width. The bins should be large enough so that the central bin is the brightest; for all sources, a bin size of $0.492''$ (1 pixel) was used. Enough bins should be used to fully enclose the source. However, diffuse source edges have proportionately more background noise, so we also constructed a larger annulus to subtract the background count rate from the source. The outer and inner radii of the background annulus were set to 5 and 1.1 times the outer radius of the source annulus, respectively.

\begin{figure}[H]
        \centering
        \includegraphics[width=0.75\linewidth]{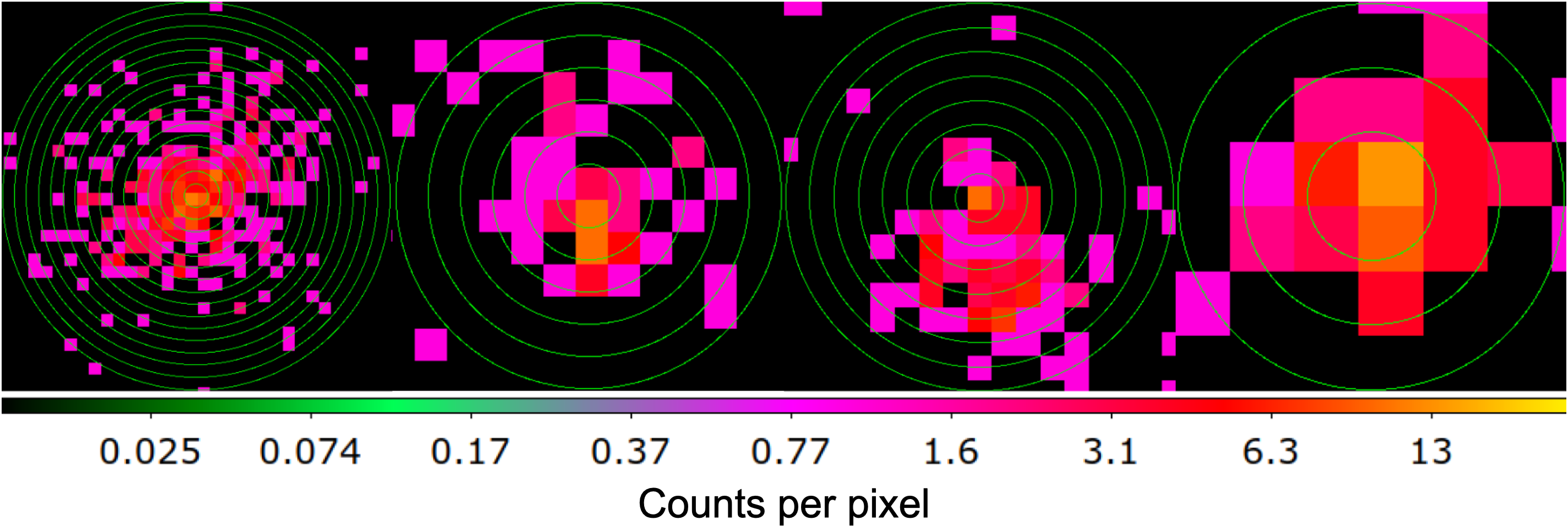}
        \caption{Original source images and radial profile annuli ($0.492''$ bin width). Left to right: XMM1 (centroid alignment, further off-axis), N1 (centroid alignment), N2 (brightest pixel alignment), N3 (centroid alignment, directly on axis).}
        \label{fig:source_images}
\end{figure}

\begin{figure}[H]
    \centering
    \includegraphics[width=0.75\linewidth]{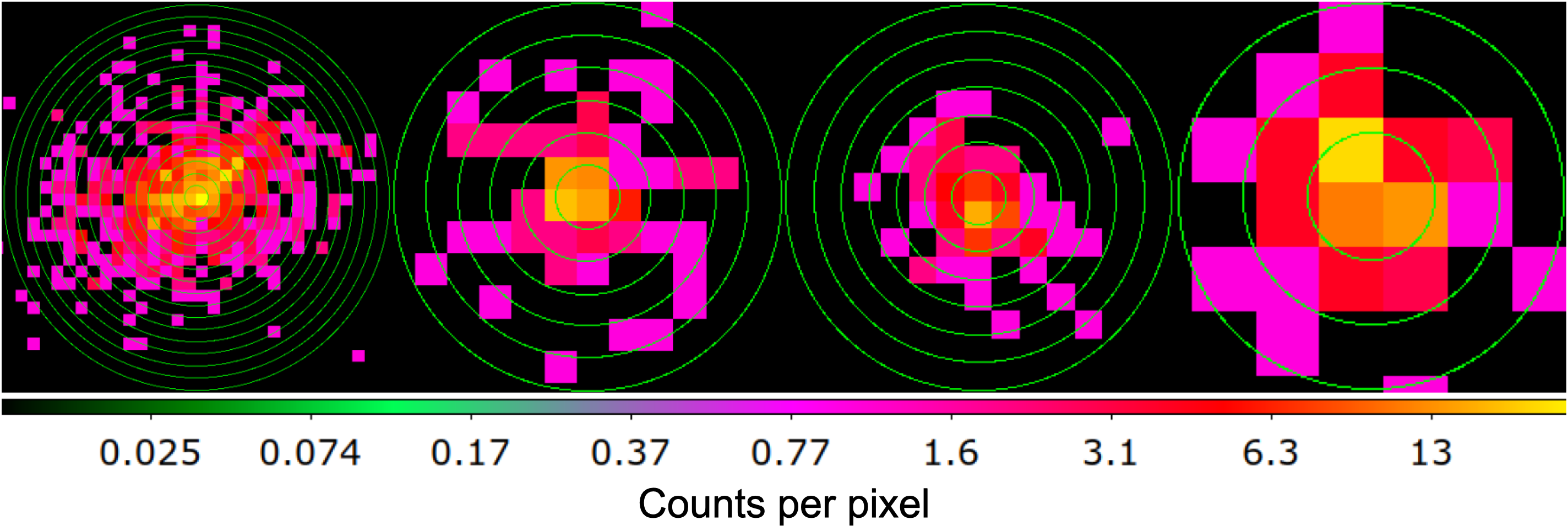}
        \caption{PSF images and radial profile annuli ($0.492''$ bin width). Left to right: XMM1 (centroid alignment), N1 (centroid alignment), N2 (centroid alignment), N3 (centroid alignment).}
    \label{fig:psf_images}
\end{figure}

Correctly aligning the annulus is also important. For symmetrical sources, the CIAO centroid function locates the mean position of all counts and aligns the annulus accordingly; for asymmetrical sources, the annulus is instead centered on the brightest pixel so that flux density decreases monotonically with radius.  For XMM1, N1, and N3, we used the centroid method; however, N2 is bright and skewed to one side, so we centered its annulus on the brightest pixel. Once the annulus is set, the CIAO command \texttt{dmextract} is used to obtain the flux density for each radial bin, for both the sources and their PSFs.

Across all observations, XMM1 contained 447 net counts at an apparent radius of $r\sim8.0''$, N1 contained 57 net counts at $r\sim3.0''$, N2 contained 113 net counts at $r\sim3.5''$, and N3 contained 57 net counts at $r\sim1.5''$; each had sufficient counts to analyze the source's true extent. Figures~\ref{fig:source_images} and~\ref{fig:psf_images} show all sources and their combined PSFs, respectively. Notably, the known point source XMM1 has a significantly more extended PSF because it lies farther off-axis from the observation center. N3 has the most concentrated PSF, at the center of its single observation. N1 was covered by three observations and was at the direct center of two of them, giving its combined PSF a slightly more extended appearance than N3. Figure~\ref{fig:merged} shows the differing off-axis positions of each source across observations. 

\begin{figure}[H]
        \centering
        \includegraphics[width=1\linewidth]{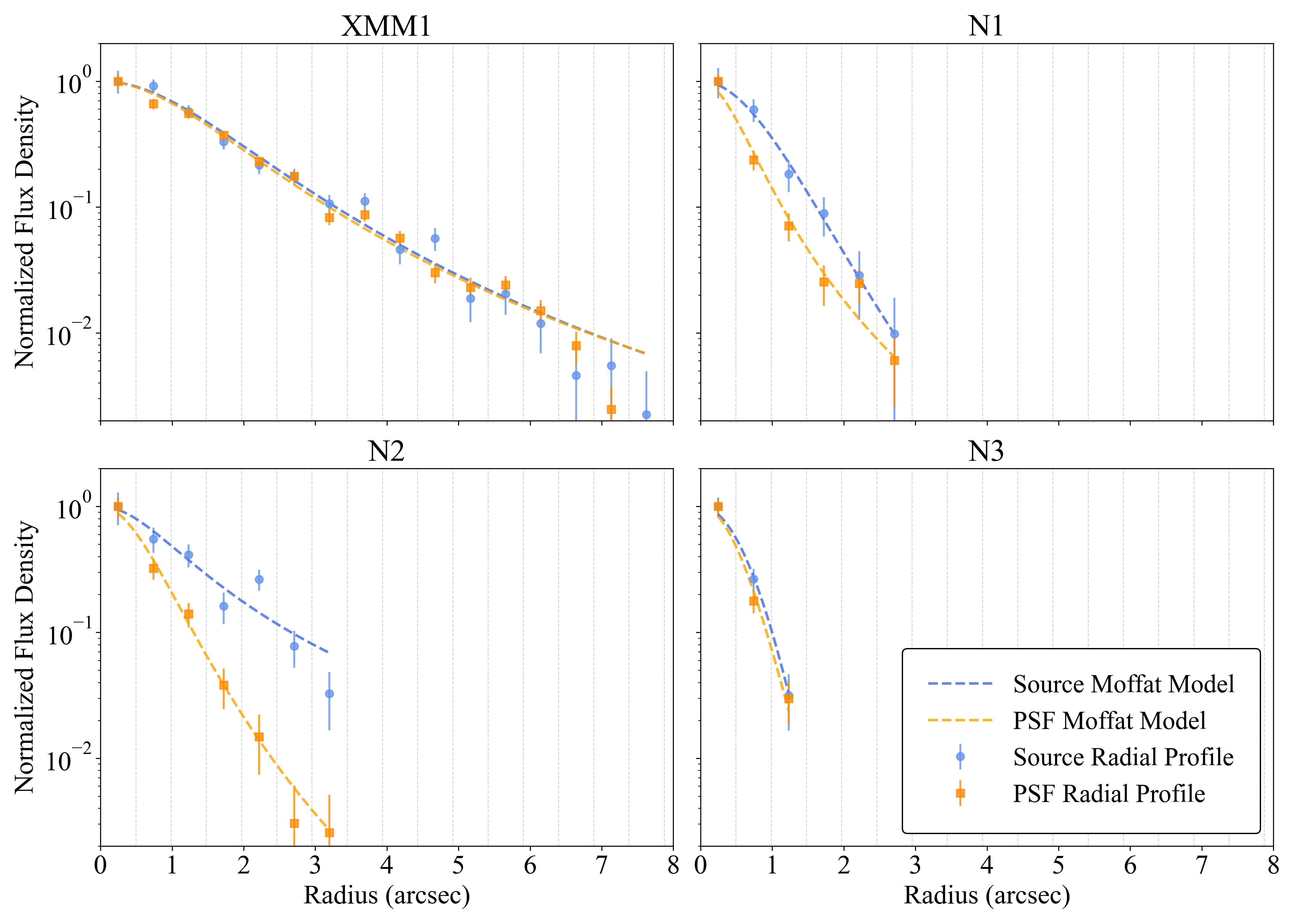}
        \caption{Source vs PSF radial profile fits modeled with the  Moffat function. Data points show the mean flux density of each radial bin ($0.492''$ width), and error bars denote standard error. Fewer overlapping error bars indicate the source is extended, and gray grid lines represent one-pixel radial intervals.}
        \label{fig:radial_fits}
    \end{figure}

With the radial profiles constructed, each source can now be directly compared to its PSF, where an extended source will show a more gradual decline in flux density with radius than its PSF does. This comparison is made in Sherpa, \chandra's plotting and fitting software \citep{Siemiginowska_2024}. In Sherpa, we first normalize each source's radial profile by dividing all bins by the central-bin flux density, then fit it with Sherpa's built-in Moffat/beta1d function:
\begin{equation}
\label{eq:moffat}
\mathrm{Flux\ density} = \mathrm{ampl}\times\,\left( 1 + \left( \frac{x - x_{\rm pos}}{r_0} \right)^2 \right)^{0.5 - 3\beta}.
\end{equation}
The Moffat function is commonly used for radial profile modeling because it parametrizes the profile's slope, core radius, position, and amplitude. In Sherpa, the $x$-position parameter is frozen at 0, and the amplitude parameter is frozen at the normalization constant. After fitting the Moffat function to both the source and PSF radial profiles, the two are plotted together in Figure~\ref{fig:radial_fits} for comparison. We also plot the standard error to quantify random deviations in flux density. Table~\ref{tab:moffat} shows the resulting best-fit parameters.

\subsection{Radial Profile Deconvolution}\label{sec:deconv}
\begin{figure}[H]
        \centering
        \includegraphics[width=1\linewidth]{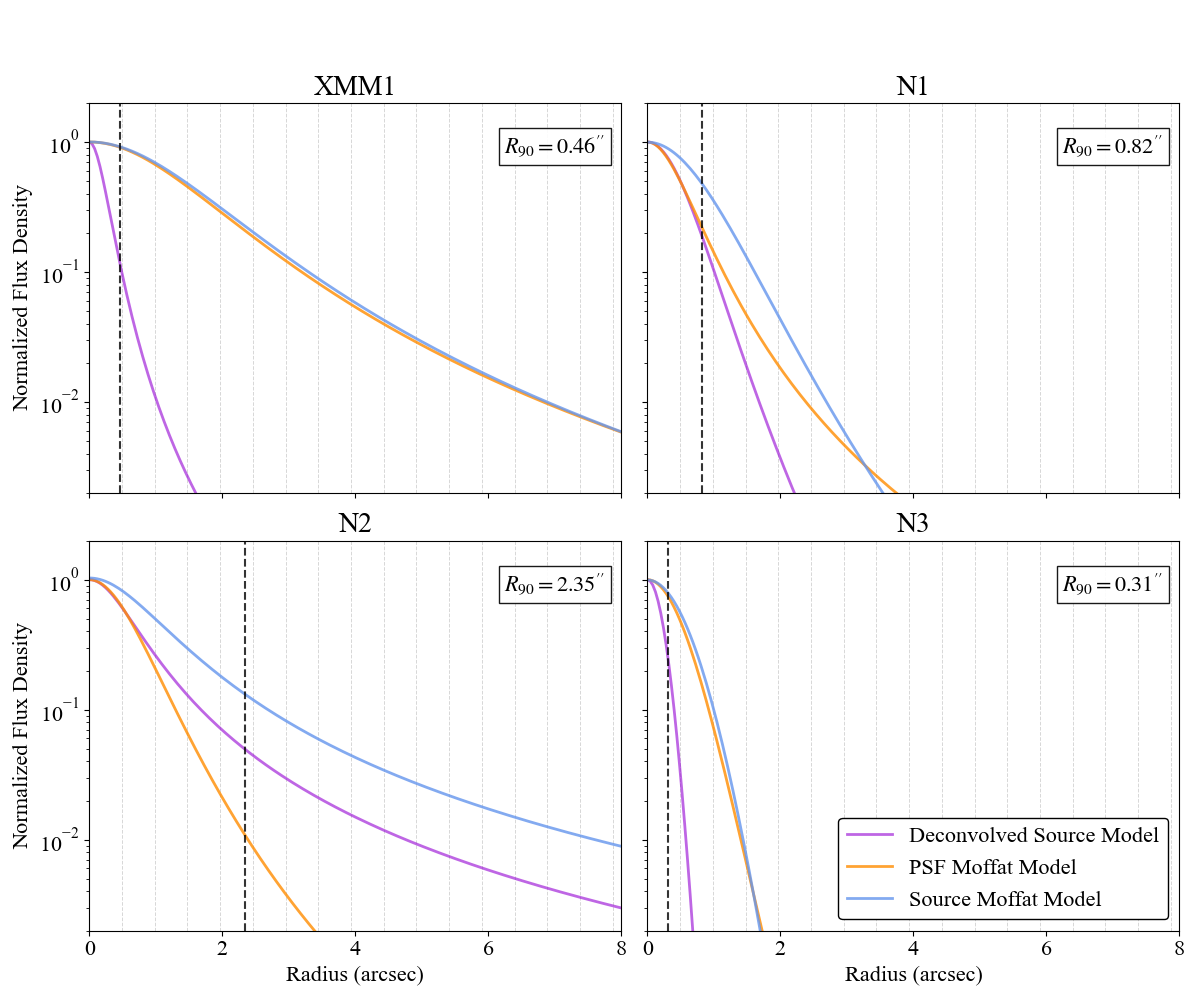}
        \caption{Radial profile deconvolution modeled with the Moffat function; the dashed vertical line marks the radius enclosing $90\%$ of the deconvolution flux density integral; the gray grid lines represent 1-pixel radial intervals.}
        \label{fig:deconv}
    \end{figure}

To recover a source's true radial extent prior to PSF distortion, the source model is deconvolved with the PSF model. We assume that a source's true (deconvolved) radial profile can also be described by Equation~(\ref{eq:moffat}). Because both $r_0$ and $\beta$ affect the shape of the flux density distribution, $r_0$ alone cannot reliably serve as a measure of true source extent. Instead, we use the radius enclosing $90\%$ of the deconvolved fit's integrated flux, hereafter $R_{90}$, as an approximate boundary for the source extent, as illustrated in Figure~\ref{fig:deconv}. The Moffat function parameters and $R_{90}$ values for the deconvolved sources are shown in Table~\ref{tab:moffat}.

\begin{deluxetable}{lccccccc}[H]
\tablecaption{Moffat Parameters and Deconvolved $R_{90}$ Values\label{tab:moffat}}
\tablehead{
\colhead{Source Name} & \colhead{$r_{\rm src}$} & \colhead{$\beta_{\rm src}$} & \colhead{$r_{\rm PSF}$} & \colhead{$\beta_{\rm PSF}$} & \colhead{$r_{\rm true}$} & \colhead{$\beta_{\rm true}$} & \colhead{$R_{90}$} \\
\colhead{} & \colhead{(arcsec)} & \colhead{} & \colhead{(arcsec)} & \colhead{} & \colhead{(arcsec)} & \colhead{} & \colhead{(arcsec)}
}
\startdata
XMM1 & 2.15 & 0.80 & 1.99 & 0.77 & 0.32 & 0.80 & 0.46 \\
N1 & 1.82 & 1.49 & 0.74 & 0.80 & 1.14 & 1.49 & 0.82 \\
N2 & 1.07 & 0.55 & 1.10 & 1.05 & 0.68 & 0.56 & 2.35 \\
N3 & 3.50 & 10.0 & 1.62 & 2.90 & 1.41 & 10.0 & 0.31 \\
\enddata
\tablecomments{Moffat parameters are given for the source, PSF, and deconvolved (true) models, together with the deconvolved $R_{90}$ value for each source.}
\end{deluxetable}

The point source XMM1 has a non-zero $R_{90}$ of $0.46''$, which is attributable to its greater off-axis angle of $6.39'$ (Table~\ref{tab:xray_sources}). The larger off-axis angle extends the PSF and increases the uncertainty of the deconvolved model, although $0.46''$ is still lower than \chandra's pixel size of $0.492''$. N3, in contrast, is on-axis and has an even smaller $R_{90}$ of $0.31''$, but at this scale it would be imprudent to claim that N3 is anything beyond a point source. Thus, XMM1 and N3 are classified as point sources, N1 as slightly extended, and N2 as clearly extended. However, given these sources' locations within a molecular cloud, and the high $N_{\rm H}$ value of N1 observed by XMM+\nustar, additional analysis is needed to determine whether the observed radial extension is instead caused by dust scattering.

\section{Dust Scattering Analysis}\label{sec:dust}
In \S\ref{sec:imaging}, we resolved the radial extent of each source prior to its entry into \chandra's optics; however, to confirm both the reliability of this technique and N2's status as an extended compact cloud clump, dust scattering must be accounted for. Dust scattering occurs when interstellar dust grains deflect light before it reaches the telescope, potentially giving a point source an artificially extended appearance \citep{2017MNRAS.468.2532J}. This effect is particularly relevant for sources within molecular clouds, since dust particle density correlates with molecular gas density; to rule out its influence in extending N1 and N2, dust scattering must be modeled.

\begin{figure}
    \centering
    \includegraphics[width=0.75\linewidth]{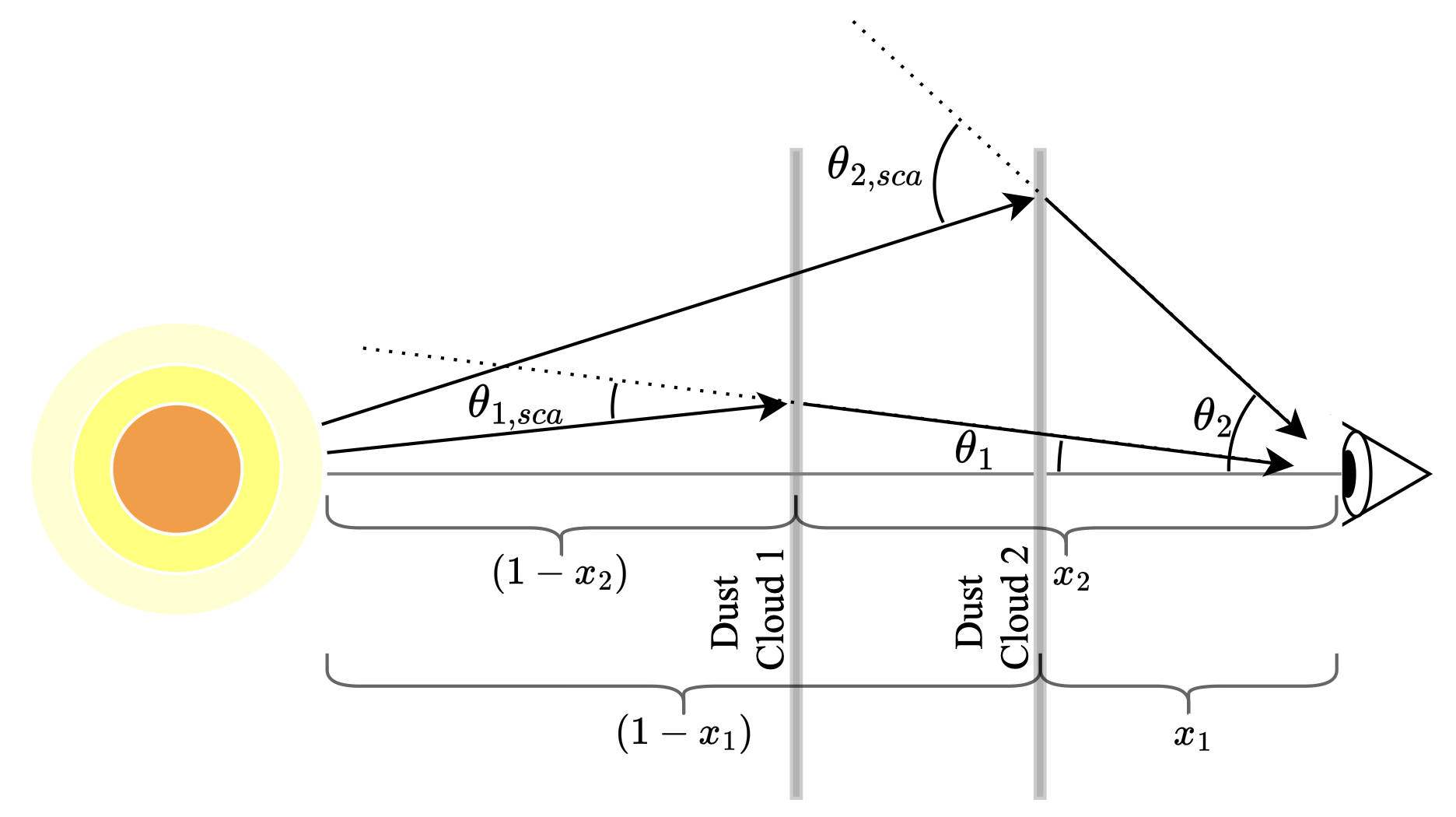}
        \caption{Diagram of dust scattering geometry. An X-ray-emitting source is on the left, and an observer is on the right. Arrows indicate the trajectories of light rays scattered toward the observer, while gray vertical lines denote dust clouds along the sight line. The parameters $x_1$ and $x_2$ represent the fractional positions of dust cloud 1 and dust cloud 2 between the source and observer. For each cloud, $\theta_1$ and $\theta_2$ are the angles to the normal of the scattered ray as seen by the observer, and  $\theta_{1,\rm sca}$ and $\theta_{2,\rm sca}$ are the corresponding scattering deflection angles.}
    \label{fig:dust_geom}
\end{figure}

Dust scattering is commonly modeled for a large dust cloud positioned between the source and observer, as illustrated in Figure~\ref{fig:dust_geom}. Strong dust scattering is associated in particular with high molecular hydrogen density, since hydrogen itself does not cause significant scattering, but molecular hydrogen forms on the surface of dust grains \citep{2017MNRAS.468.2532J}. Toward G22.7$-$0.2, the line-of-sight $N_{\rm H_2} \sim 3.7 \times 10^{22}\,\mathrm{cm}^{-2}$, which is attributed to GMC~23.3$-$0.4  \citep{2026ApJ..1001...19M}.  This is a considerably high $N_{\rm H_2}$ value, indicating that GMC~23.3$-$0.4  could be a significant contributor to dust scattering for N1, N2, and N3. Any remaining column density can likely be attributed to atomic hydrogen in the ISM and in GMC~23.3$-$0.4, both of which cause negligible dust scattering for hard X-ray sources \citep{2017MNRAS.468.2532J}.

Dust scattering flux intensity for a point source separated from the observer by a dust cloud, $I_{\mathrm{sca}}(\theta)$, is modeled by Equation~(\ref{eq:scat}) \citep{1991ApJ...376..490M}.

\begin{equation}
\label{eq:scat}
I_{\mathrm{sca}}(\theta)
=
F_X N_{\rm H}
\int_{E_{\min}}^{E_{\max}}
S(E)
\int_{0}^{1}
\frac{f(x)}{(1-x)^2}
\int_{a_{\min}}^{a_{\max}}
n(a)\,
\frac{d\sigma_{\mathrm{sca}}(a,x,E,\theta)}{d\Omega}
\, da\, dx\, dE.
\end{equation}

The parameters in Equation~(\ref{eq:scat}) are: $F_X$, the X-ray flux; $N_{\rm H}$, the hydrogen column density; $a$, the dust grain size; $x$, the fractional position of the dust cloud between the observer and source; $E$, the emitted energy of the source; and $\theta$, the offset angle from the source. Functions $n(a), f(x),$ and $ S(E)$ are the normalized distributions of these parameters. 

The term $\frac{d\sigma_{\mathrm{sca}}(a,x,E,\theta)}{d\Omega}$, the differential scattering cross section of a single dust grain (cross section per unit solid angle), is given by:

\begin{equation}
\label{eq:dsigma}
\frac{d\sigma_{\mathrm{sca}}}{d\Omega} = \left(1.1\ \mathrm{cm}^2\,\mathrm{sr}^{-1}\right)
\left(\frac{2Z}{M}\right)^2
\left(\frac{\rho}{3\ \mathrm{g\,cm}^{-3}}\right)
\left(\frac{F(E)}{Z}\right)^2
\times
\left(\frac{a}{1\ \mu\mathrm{m}}\right)^6
\, \exp\left[
-0.4575\, E(\mathrm{keV})^2\, a(\mu\mathrm{m})^2\, \theta_{\mathrm{sca}}(\mathrm{arcmin})^2
\right].
\end{equation}

Here, $\rho$ is the dust density, $Z$ is the mean atomic number of the dust, $M$ is the mean atomic mass of the dust, and $F(E)$ is the atomic scattering factor. The dust scattering angle $\theta_{\mathrm{sca}}$ is given by Equation~(\ref{eq:thetasca}), which follows from the observer--dust--source geometry (Figure~\ref{fig:dust_geom}) for a single scattering \citep{2017MNRAS.468.2532J}. Because this holds for a grain at any $x$, it can be integrated over $x$ to account for a dust cloud of finite thickness.

\begin{equation}
\label{eq:thetasca}
\theta_{\mathrm{sca}} = \arctan\left(\frac{x}{1-x}\tan\theta\right) + \theta.
\end{equation}

Dust scattering could only lead to misidentification of N1 or N2 if it produced proportionately high flux at  $\theta\sim 0.5''$--$3.5''$; it is therefore sufficient to model the relative dust scattering intensity as a function of $\theta$, rather than model every dust scattering parameter. Taking the average width of GMC~23.3$-$0.4 to be $15 \pm 2$~pc, from the cloud mass and mean density \citep{2026ApJ..1001...19M}, and the distance to G22.7$-$0.2 of $4.55\pm0.15$~kpc \citep{Su_2014, Ranasinghe_2018}, $x$ is bounded by $0.996<x<1$. Assuming that the scattering dust is primarily concentrated in a single layer whose thickness equals the width of GMC~23.3$-$0.4, Equation~(\ref{eq:scat}) simplifies to:

\begin{equation}
\label{eq:iscat}
I_{\mathrm{sca}}
\propto
\int_{0.996}^{1}
\frac{1}{(1-x)^2}\, \exp\left[
-\left(\arctan\left(\frac{x}{1-x}\tan\theta\right)\,(\mathrm{arcmin}) + \theta\,(\mathrm{arcmin})\right)^2
\right]
\,dx.
\end{equation}

\begin{figure}
    \centering
    \includegraphics[width=0.75\linewidth]{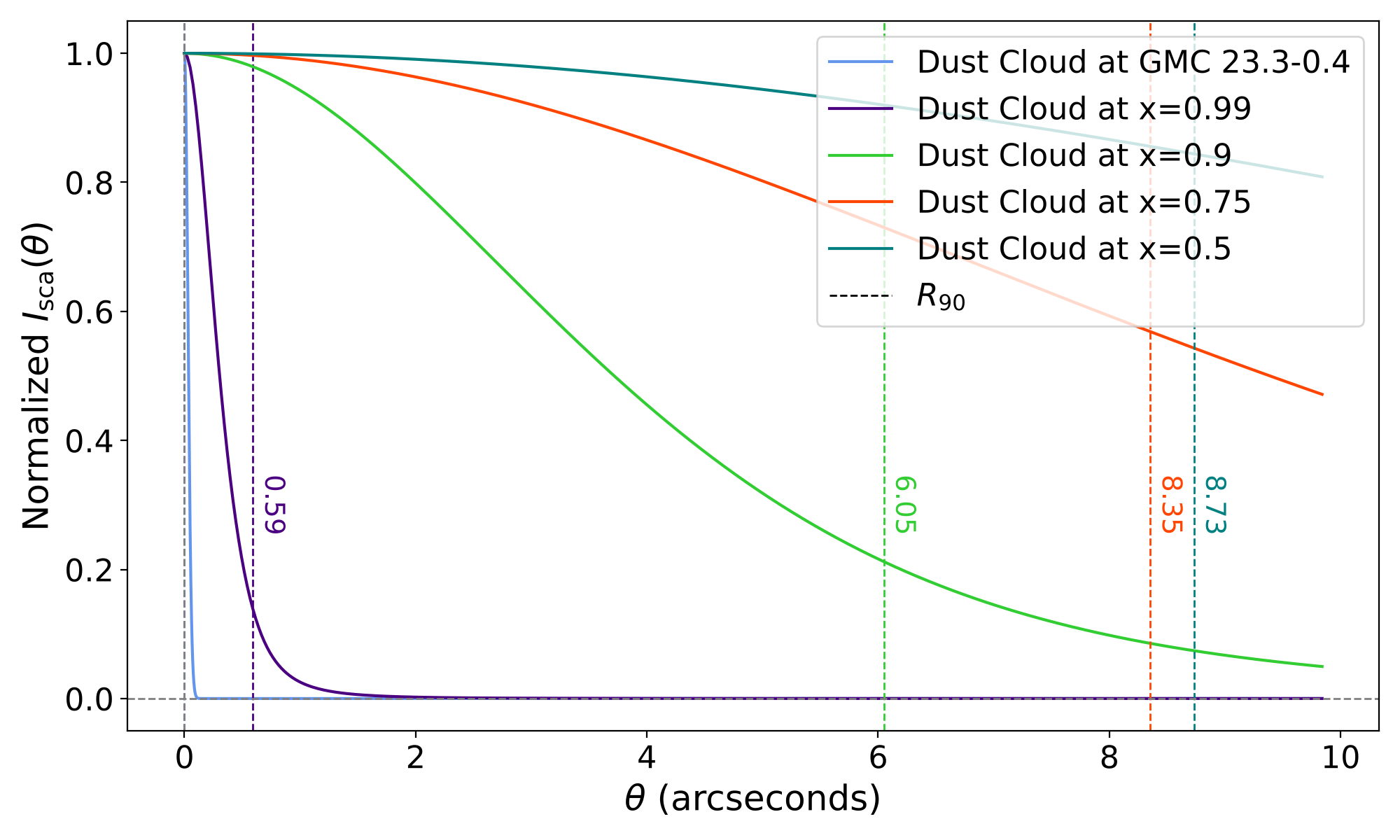}
    \caption{Normalized dust scattering intensity versus observed angle for dust clouds at different positions. The blue line represents GMC~23.3$-$0.4, where the dust clouds are located directly at the source ($x=1$). All models assume a fractional cloud width of $0.004$, matching that of GMC~23.3$-$0.4. The colored dashed lines represent the $R_{90}$ values in arcseconds from each dust cloud model.}
    \label{fig:dust_prof}
\end{figure}

For direct comparison with the radial profiles, Figure~\ref{fig:dust_prof} plots this proportional intensity against radial offset angle.

Figure~\ref{fig:dust_prof} shows that, for a source embedded within a distant dust cloud, $I_{\mathrm{sca}}(\theta)$ behaves effectively as a $\delta$ function, so point sources will still appear as points. N1 and N2 both have $R_{90}$ values greater than that of a cloud at $x=0.99$, so this holds as long as there is no evidence for a large, dense dust cloud $\gtrsim50$~pc in front of GMC~23.3$-$0.4.

In \citet{Su_2014} the distance to GMC~23.3$-$0.4 is derived from its line-of-sight velocity, which a Galactic rotation curve converts to a kinematic distance of $\sim4.4$\,kpc, and \citet{2026ApJ..1001...19M} adopt this distance. Toward $l\sim 22.6^\circ,b=-0.1^\circ$, in front of N1 and N2, intensity clearly peaks at $v_{\rm LSR}\sim75$\,km\,s$^{-1}$ indicating a single cloud \citep{2026ApJ..1001...19M}. \citet{Su_2014} find a $\sim$60--70\,km\,s$^{-1}$ CO component in the south-east of the remnant ($l\approx22.8^\circ$, $b\approx-0.5^\circ$) that may indicate foreground contamination elsewhere in the system, but neither paper reports additional foreground molecular gas at $(22.6^\circ,-0.1^\circ)$ in front of N1 and N2.

Moreover, a dense, extended cloud in front of GMC 23.3$-$0.4 would likely affect multiple sources, not just one or two. In Appendix~\ref{app:psf}, each source in Table~\ref{tab:xray_sources} is shown next to its PSF. None of the 23 sources other than N1 and N2 are clearly extended beyond their PSF shape. If those 23 sources are all point-like, it is very unlikely that only the clump candidates N1 and N2 happen to lie behind a large, dense cloud.

\section{Results and Discussion}\label{sec:results}
The G22.7$-$0.2 SNR region is a rich case study of SNR--MC interactions, so this work began with surveying and cataloging the X-ray sources within it. We then aimed to resolve N1, N2, or N3 as extended hard X-ray sources, since extension could be explained only by shock-heated CCCs emitting either synchrotron or NTB radiation. Both NTB-emitting CCCs and CCCs in a middle-aged SNR are novel observational phenomena, making their discovery significant and creating broader scientific implications. 

\subsection{Characterizing the CCC Candidates Discovered in the \chandra\ Survey}\label{sec:cand}
In \S\ref{sec:deconv}, the true source radii were estimated at $r\sim0.82''$ for N1, and $r\sim2.35''$ for N2. XMM1 is already a known point source resolved to $r\sim0.46''$, and N3 is even smaller at $r\sim0.31''$, so we take both to be point sources. However, for N1 and N2, these angular extents can be converted to approximate physical length scales given the known distance range of 4.4--4.7~kpc: N1 has $r\sim0.018\pm0.0006$~pc and N2 has $r\sim0.052\pm0.002$~pc. 

Since the radius alone does not constrain the mass, we use the densities implied by the two candidate emission models, together with the mean nucleon density of the cloud measured by \citet{2026ApJ..1001...19M} at $n_{\rm p}\sim5.3\times10^{2}\,{\rm cm^{-3}}$. For the slow-shock slab model from \citet{Bykov_2000}, $n_{\rm H}\sim3\times10^{4}\,{\rm cm^{-3}}$, a uniform sphere has $M\approx0.03\,M_\odot$ for N1 and $M\approx0.6\,M_\odot$ for N2, whereas for the $10^{5}$--$10^{7}\,{\rm cm^{-3}}$ cores of the synchrotron hotspot model \citep{2020ApJ...899..102H} N1 has $M\approx0.08$--$8\,M_\odot$ and N2 has $M\approx2$--$200\,M_\odot$. The slab masses lie below the empirical mass limit for massive-star formation at these length scales \citep{2010ApJ...716..433K}, whereas the synchrotron-regime masses exceed it for densities of $n_{\rm H}\gtrsim5\times10^{6}$ and $8\times10^{5}\,{\rm cm^{-3}}$ for N1 and N2, respectively, placing the clumps in the massive-star-forming regime. The slab density is only $\sim60$ times the mean cloud density, compared with $\gtrsim200$ times for the synchrotron cores, which makes the shocked slab the more plausible interpretation, at least for N2. It should be noted that, as the images in Figure~\ref{fig:source_images} show N2 to be an asymmetric source, the mass and density estimates, which assume spherical symmetry, should be treated as approximations only.

N1 ($r\sim0.018$~pc) and N2 ($r\sim0.052$~pc) were the only sources resolved as extended, and both lie within the length scale of a CCC ($r\sim0.01$--$0.1$~pc). N1 has a softer photon index of $\Gamma= 2.1^{+0.5}_{-0.4}$ \citep{Mori_2017}, more characteristic of a synchrotron-emitting hotspot \citep{Kawabata2026upstreamB}. This is plausible for N1 as its smaller clump scale ($r\sim0.018$~pc) correlates with a higher gas density, which is needed for synchrotron emission. Alternatively, \citet{Zhang_2014} detected a non-thermal X-ray filament near a SNR--MC interaction site with spectral properties similar to N1 ($\Gamma= 2.28^{+0.17}_{-0.18}$, $N_{\rm H} > 10^{23}\,\mathrm{cm}^{-2}$), but attributed it to a magnetic flux tube in a locally amplified B-field. N2 has a photon index of $\Gamma= 1.2^{+0.4}_{-0.4}$ \citep{Mori_2017}, consistent with the $\Gamma\simeq1$ expected for NTB emission from a loss-flattened electron distribution \citep{Uchiyama_2002, 10.1111/j.1365-2966.2009.14832.x}, which assumes gas dense enough for Coulomb losses to flatten the electron spectrum at the energies that produce the X-rays, an assumption that is plausible for these sources \citep{Uchiyama_2002}. Such a hard photon index in an extended source of $r\sim0.01$--$0.1$~pc is most plausibly explained by a CCC emitting NTB radiation from LECRs cooling within the SNR shell \citep{Bykov_2000}. This interpretation is further supported by N2's morphology, as its image in Figure~\ref{fig:source_images} is visibly skewed toward the southeast, approximately aligned with the expanding shockwave of G22.7$-$0.2. Beyond the $\Gamma\simeq1$ NTB signature, synchrotron emission could be distinguished by a radio counterpart or by X-ray polarimetry \citep{Dubner_2015}. 20~cm MAGPIS observations \citep{Helfand_2006} detected no emission from N1, N2, or N3 above the surrounding background, but synchrotron emission in the region could be tested with follow up polarimetry observations.

The gas ionization rate and shock compression ratio of CCCs are also comparable to those of typical molecular gas along a SNR shell, indicating that while CCCs are unlikely to be fully ionized, their surfaces may still become ionized, with penetration depths $\gtrsim1\%$ well short of the column density through a full clump \citep{Bykov_2000}. This surface ionization could help explain why N1 and N2 sit toward the small end of the CCC size range, if larger clumps have simply had more of their outer layers eroded away over time. The shock interaction is also expected to compress a clump into a dense slab, with the shock-facing side compressed and processed well before the leeward side catches up. This can cause a clump's shape to distort asymmetrically along the shock front, as appears to be the case for N2.

\subsection{Broader Implications and Future Directions of Non-thermal Bremsstrahlung Radiation and CCCs}\label{sec:implications}
NTB radiation illuminating dense MC substructures on a SNR shock wave was first predicted over 25 years ago, but observational evidence has been lacking until now \citep{Bykov_2000}. Previous observations of non-thermal CCCs have been dominated by synchrotron-emission models, as in the case of RX J1713.7$-$3946 \citep{2020ApJ...899..102H,Kawabata2026upstreamB}. \nustar\ observations of the SNR IC 443 have found hotspots with similar NTB-emitting CCC characteristics, but these could not be resolved without \chandra's angular resolution \citep{Zhang2018NuSTARDO}. The X-ray-luminous MC substructures presented here represent a distinct class, observable only thanks to \chandra's sub-arcsecond resolution. This new regime for studying SNR--MC interactions creates new possibilities for future research, both for G22.7$-$0.2 itself and for the wider population of SNRs known to interact with molecular gas.

Observations of NTB-emitting MC substructures can be combined with other tracers to build more accurate models of MC and ISM structure, particularly in cases where molecular-line and sub-mm dust-emission observations lack sufficient spatial resolution. Other CCC observations could also be used to test the prediction by \citet{Bykov_2000} that sufficiently fast SNR shocks can drive CCCs to emit soft $\gamma$-rays; N2 itself is unlikely to be a useful test case, however, due to contamination from the nearby $\gamma$-ray emitting source HESS J1832$-$093. NTB-emitting CCCs have also been modeled as variable over time ($\sim$few years) \citep{Bykov_2000}, unlike synchrotron-emitting CCCs, which have been observationally stable over 10 years \citep{Kawabata2026upstreamB}. Thus, multi-epoch Chandra observations of N2 and other non-thermal CCCs could be another method of differentiating NTB from synchrotron emission, as well as studying MC substructure evolution from SNR shocks.

Finally, the method used here to resolve N1 and N2 is not specific to these G22.7$-$0.2 observations. CCCs may be a common but overlooked source population elsewhere, mistaken for other source types due to their subtle radial extent. Applying the same techniques to archival \chandra\ data from other well-studied SNR--MC systems, such as IC 443, W44, and RX J1713.7$-$3946, could extend the study of CCCs from a single candidate to a broader population.

\section{Conclusion}\label{sec:conclusion}
The primary goal of this study was to exploit \chandra's high angular resolution to determine whether three hard X-ray sources detected by \nustar\ and other X-ray sources in the shell of SNR G22.7$-$0.2 are extended, representing CCCs emitting non-thermal X-rays. We modeled each source's PSF and compared it to the source's radial profile, using the known point source XMM1 as a control. We found that dust scattering within the molecular cloud contributed negligibly to the observed source extent. This analysis identified N3 as a point source, N1 as a possible CCC, MC filament, or synchrotron-emitting clump with $r\sim0.018$~pc, and N2 as a compelling CCC candidate with $r\sim0.052$~pc and $M\approx0.6\,M_\odot$ under the NTB model, thus lending compelling support to the theory of NTB-emitting CCCs. The secondary goal of this work was to further catalog the scientifically rich G22.7$-$0.2 region. Using the CIAO tool {\tt wavdetect}, we identified 23 sources with significance $>5\sigma$, of which 20 are newly cataloged X-ray sources. Similar \chandra\ X-ray surveys of SNRs interacting with molecular clouds could reveal CCCs emitting non-thermal X-rays in the future.

\section*{Acknowledgments}
We express our gratitude to Dr.\ Patrick Slane, Dr.\ Shifra Mandel, and Dr.\ Rafael Martinez-Galarza for their scientific insight, and guidance on the \chandra\ analysis presented in this work. We also thank the \chandra\ X-ray Center and its staff for providing high-resolution data, maintaining thorough documentation, and hosting annual workshops that greatly facilitated our analysis. This work was supported by the \chandra\ grant SAO GO2-23021X.

\facilities{CXO, 2MASS Point Source Catalog}

\software{CIAO v4.17 \citep{2006SPIE.6270E..1VF}, CALDB v4.12.4, Sherpa \citep{Siemiginowska_2024} v4.17, wavdetect \citep{Freeman_2002}, ChaRT v2 \citep{2003ASPC..295..477C}, MARX v5.5.3 \citep{2012SPIE.8443E..1AD}, PIMMS \citep{Mukai1993PIMMS}}

\appendix
\section{Source Images versus their PSF}\label{app:psf}
\begin{figure}[H]
    \centering
    \includegraphics[width=0.9\linewidth]{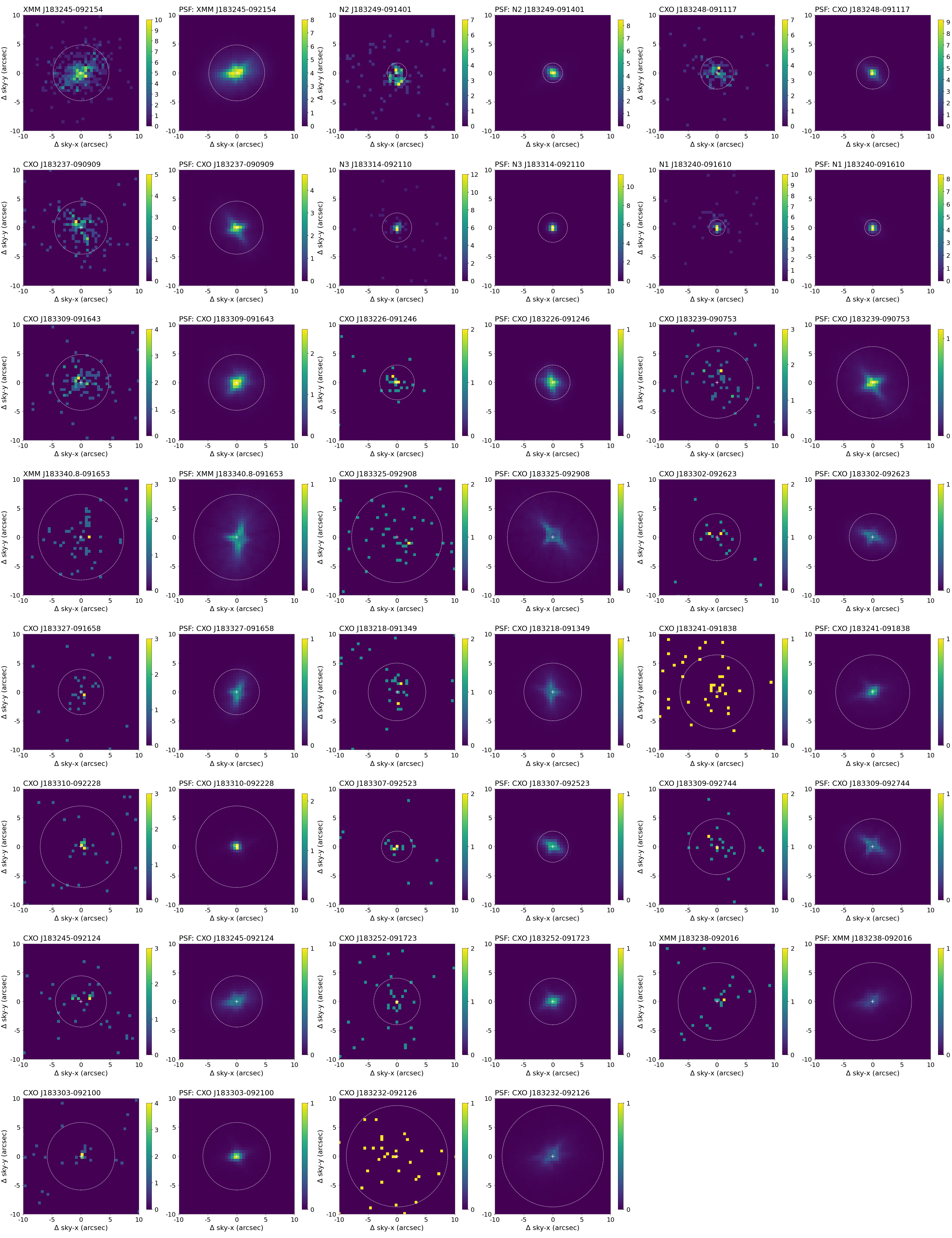}
    \caption{Image of each source from Table~\ref{tab:xray_sources}, with its PSF shown to the left. Each image is a 20 by 20 arcsecond cutout, and the dashed ring marks the approximate 90\% enclosed flux radius of the corresponding PSF. The color bar gives the number of counts per pixel.}
    \label{fig:figure8}
\end{figure}

To support the assertion that N1 and N2 are the only sources in Table~\ref{tab:xray_sources} that are clearly extended relative to their PSFs, we present the image and PSF of each source in Figure~\ref{fig:figure8}. For many of these sources, simulating the PSF with ChaRT goes beyond what is necessary, so we instead use MARX \citep{2012SPIE.8443E..1AD} with a monoenergetic spectrum at the logarithmic mean of the 0.5--7 keV band. Otherwise, each PSF is processed as in \S\ref{sec:psf}, and the 90\% enclosed flux radius is also shown for each PSF. N1 and N2 stand out in two ways: their morphologies differ clearly from their PSFs, and a large fraction of their flux falls outside the 90\% enclosed-flux radius. Both also lie near the optical axis, where the PSF is most compact.

\raggedbottom

\flushbottom
\bibliography{references}

\end{document}